%% file: main.tex
\documentclass{ieeeaccess}
\usepackage{cite}
\usepackage{amsmath,amssymb,amsfonts}
\usepackage{algorithmic}
\usepackage{graphicx}
\usepackage{textcomp}
\usepackage{url}
\usepackage{stfloats}
\usepackage{hologo}
\usepackage{listings}
\def\BibTeX{{\rm B\kern-.05em{\sc i\kern-.025em b}\kern-.08em
    T\kern-.1667em\lower.7ex\hbox{E}\kern-.125emX}}
\begin{document}
\input{copyright} 
\history{Received 10 October 2022, date of current version 10 October 2022.}
\doi{10.1109/ACCESS.2022.DOI}

\title{Behavior-Semantic Scenery Description (BSSD) of Road Networks for Automated~Driving}
\author{\uppercase{Moritz Lippert}\authorrefmark{1}, \uppercase{Felix Glatzki}\authorrefmark{1}, and \uppercase{Hermann Winner}\authorrefmark{2}}
\address[1]{Institute of Automotive Engineering, Technical University of Darmstadt, 64287 Darmstadt, Germany}
\address[2]{Former professor for Automotive Engineering at Technical University of Darmstadt, 64287 Darmstadt, Germany}
\tfootnote{This research is supported by the Federal Ministry of Education and Research of Germany (BMBF) within the project UNICAR\textit{agil}~(FKZ~16EMO0286) and by Continental within the project PRORETA 5.\\ Authors M. Lippert and F. Glatzki contributed
equally to this work.}

\markboth
{Lippert \headeretal: Behavior-Semantic Scenery Description (BSSD) of Road Networks for Automated Driving}
{Lippert \headeretal: Behavior-Semantic Scenery Description (BSSD) of Road Networks for Automated Driving}

\corresp{Corresponding author: Moritz Lippert (e-mail: moritz.lippert@tu-darmstadt.de).}

\begin{abstract}
The safety approval of Highly Automated Vehicles (HAV) is economically infeasible with current approaches. For verification and validation, it is essential to describe the intended behavior of an HAV in the development process in order to prove safety. The demand for this behavior comes from the traffic rules which are instantiated by the present scenery around the vehicle (e.g. traffic signs or road markings). The Operational Design Domain (ODD) specifies the scenery in which an HAV may operate, but current descriptions fail to explicitly represent the associated behavioral demand of the scenery. We propose a new approach for a Behavior-Semantic Scenery Description (BSSD) in order to describe the behavior space of a present scenery. A behavior space represents the delimitation of the legally possible behavior. The BSSD explicitly links the scenery with the behavioral demand for HAV. Based on identified goals and challenges for such an approach, we derive requirements for a generic structure of the description for complete road networks. All required elements to represent the behavior space of the scenery are identified. Within real world examples, we present an instance of the BSSD integrated into the HD-map framework Lanelet2 to prove the applicability of the description. The presented approach supports development, test and operation of HAV by closing the knowledge gap of where a vehicle has to behave in which limits within an ODD.
\end{abstract}

\begin{keywords}
Automated vehicles, behavioral requirements, operational design domain, scenery description, vehicle safety
\end{keywords}

\titlepgskip=-15pt

\maketitle

\section{INTRODUCTION}
\label{sec:introduction}
\PARstart{A}{utomated} vehicles are currently the focus of research and development both in the automotive industry and in vehicle technology research institutions. The need for a verification and validation as well as a safety by design solution of these automated systems becomes clear not least in the recently published technical report ISO/TR 4804 \cite{ISO.2020}. However, in order to be able to develop, test and release the functions required for a driving automation system under these aspects, the functions must first be explicitly specified. This specification is fundamentally carried out in the description of the Operational Design Domain (ODD), in which the operational area of an automated vehicle is defined \cite{SAEInternational.2014}. The specification of the operational area indirectly defines the possible interactions of the vehicle with its environment within this area. In this context, the vehicle environment consists of the static traffic environment such as roads or further traffic infrastructure as well as other road users or other objects. All interactions of a Highly Automated Vehicle (HAV) with its environment can be understood as vehicle behavior. Consequently, by defining this vehicle environment, the ODD indirectly specifies the behavior rules and thus the behavior limits for HAV. In order to explicitly represent this indirect information and therefore clearly define the possible vehicle behavior, a crucial question must first be answered: What behavioral rules and limits are imposed on the vehicle by an ODD?

Depending on the country, traffic regulations describe the applicable traffic rules in general terms (e.g., German Road Traffic Regulations \cite{BundesministeriumderJustizundfurdenVerbraucherschutz.2013}). There are global rules that apply everywhere regardless of the scenery (e.g. collision-free driving as prominently addressed in \cite{ShalevShwartz.8212017}) and local rules that only arise in combination with concrete sceneries. The local traffic rules instantiated by the present scenery represent the previously mentioned behavioral rules and limits for all road users and thus also for HAV. Although these local rules are not directly concerned with collision avoidance, these rules are highly relevant to road safety. Even if, for example, collisions are avoided at intersections, subsequent accidents involving other traffic participants can still occur due to a disregard of the applicable local behavioral rules. This shows that not only collision avoidance itself contributes to road safety, but also the aforementioned local rules, which are therefore indispensable. So far, the unified derivation and linkage of these local traffic rules based on the scenery has not been addressed in the literature. Therefore, the focus of this paper is on local traffic rules and the resulting behavioral limits. Only the legally relevant behavior limits are considered and explicitly no other behavior limiting factors such as visibility conditions or reduced friction values due to weather or other influences.

The identified lack of process results in unresolved challenges for the development process of HAV. As shown before, the necessary driving behavior of HAV is based on the ODD. ISO 21448 \cite{SOTIF} gives guidance on how to ensure that the intended functionality is safe. This means that there is absence of unreasonable risk due to hazards resulting from functional insufficiencies of the intended functionality or its implementation. This includes insufficiency of specification \cite{SOTIF}. A complete and explicit representation of the constraints on vehicle behavior imposed by traffic rules is therefore indispensable within the functional specification. The functional specification forms the basis for the systematic derivation of requirements for the driving automation in the system development process. However, these requirements are only valid for the considered functional specification, so that behavioral demands not listed in the functional specification cannot be addressed. The resulting specification gap propagates to the test case definition - even with complete test coverage with regards to the previously derived requirements. A uniform and holistic description of the behavioral demands resulting from the ODD is therefore required, which needs to be available at the beginning of the development process.

Unsolved challenges also arise for the operation of HAVs. For traffic rule-compliant behavior, the surrounding scenery must be permanently analyzed while driving using a database that translates the occurring scenery elements and their concrete combinations into the applicable behavior rules. Another problem is mission planning and continuation when driving capabilities are degraded. It is not known for which road sections within the ODD which driving capabilities are required. Thus, the driving mission must be aborted in case of degradation, since otherwise no safe operation of the vehicle can be guaranteed. Thus, no route planning is possible that takes into account the current driving capabilities of HAV and checks whether they meet the behavioral requirements of a selected route. In both cases, it would be easier to store the applicable behavioral demands directly in a map, for example. A first approach to this is the map framework \textit{Lanelet2} \cite{Poggenhans.2018} which represents traffic rules within the map. However, behavioral demands are not mapped uniformly and completely.

Due to the identified lack of a behavior-related basis for the development and operation of HAV, we would like to answer the following research question within this work: How can the behavioral demands resulting from the scenery be described uniformly and linked to the scenery directly?

As a result, we introduce a universal and explicit representation of the behavioral demands directly linked to the scenery. We show that current research and development do not provide an approach for identifying and directly linking behavioral demands based on a scenery for HAV. Furthermore, we present the Behavior-Semantic Scenery Description (BSSD) of road networks for the description and characterization of arbitrary ODDs in the context of highly automated driving. This novel description provides access to understanding and knowledge of what behavioral constraints need to be fulfilled by HAV and where within a present scenery this needs to be accomplished. This approach potentially creates a unified tool base for development, verification and validation of HAV and could additionally support HAV operation.

In the following, the fundamentals and goals of this work are first explained and requirements for the BSSD are derived. Based on this, related work is analyzed with respect to already existing approaches. Subsequently, our solution of the BSSD is presented including an application example. The paper closes with conclusion and outlook.

\section{PRELIMINARIES} \label{preliminaries}
\subsection{TERMINOLOGY}
To fit this work into the understanding of the community, the two main terms \textit{scenery} and \textit{behavior} are used according to the current state of the art.
In the course of creating a scenery catalog, Geyer \textit{et al.} \cite{Geyer.2014} define the term scenery as a structured collection of individual static elements that form the environment for dynamic elements. Ulbrich \textit{et al.} \cite{Ulbrich.2015} adopt and concretize this definition of the term by describing the scenery as a summary of all geo-spatially stationary elements. In addition to stationary elements themselves, this definition also includes the lane network, vertical elevation, and environmental conditions. On top of scenery, the same paper defines the terms scene, situation, and scenario, which have since become successfully established in the automated driving community. For this reason, these terms are used in this paper according to the understanding of Ulbrich \textit{et al.} \cite{Ulbrich.2015}.

To identify and describe behavioral demands, it is first necessary to define the behavior of an HAV. According to Nolte \textit{et al.} \cite{Nolte.2017}, a distinction can be made between internal and external behavior, each of which is described by a sequence of internal and external states. Consequently, internal behavior describes in-vehicle processes to fulfill specified vehicle functions, while external behavior represents observable vehicle actions and interactions with the environment. Czarnecki \cite{Czarnecki.2018b} proposes a similar behavior definition in terms of a road user behavior. He describes this behavior as a temporal change in states caused by internal or external factors. Road user states are distinguished into internally and externally observable states. Externally observable states are the basic state of motion, physical form, and the relationship between road users and other objects. This includes, among other things, the activities of road users. Czarnecki's \cite{Czarnecki.2018b} behavior related to externally observable state changes thus fits into Nolte \textit{et al.}'s \cite{Nolte.2017} definition of external behavior. If behavior is used in this paper, the composition of these two definitions is meant. Internal processes or changes of state that cannot be observed externally are not relevant in the context of the presented approach.

\subsection{BEHAVIOR SPACE AND BEHAVIORAL ATTRIBUTES}
In a previous paper \cite{Glatzki.91920219222021}, this author team introduced the term \textit{behavior space} as the delimited set of possible behaviors for a vehicle based on traffic rules. The behavior space does not demand explicitly required behavior, but rather spans the limits of the legally allowed behavior. We call the information regarding these limits the \textit{behavioral demand}. We derived a structure of \textit{behavioral attributes} in order to describe the behavior space. This assigns the behavior relevant information to a section of the scenery. This abstraction process reduces the complexity of the scenery description while preserving the behavior relevant context and thus, is beneficial for functional specification as well as selection of the ODD. We introduced the term \textit{regular motion space} that describes the motion space for motor vehicles that is usually the roadway. On top of that, we defined four behavioral attributes with underlying properties to describe the behavior space of sections of a road network. These sections are called \textit{atomic behavior spaces}, because within them, per definition, the behavioral demand does not change.

The \textit{speed attribute} describes the maximum allowed or minimum required speed of a scenery section.  Besides the speed value itself, possible time restrictions or other conditions can be allocated.

The \textit{boundary attribute} specifies the rules and restrictions for crossing boundaries. It is differentiated between longitudinal and lateral boundaries. Crossing these boundaries may be \textit{allowed}, \textit{conditional} (= allowed under a certain condition), \textit{prohibited} or \textit{not possible} (physically).

The \textit{reservation attribute} defines conditions to enter and remain in the atomic behavior space with respect to priority rules. Every atomic behavior space is, with some exceptions, reserved for at least one type of traffic participant (e.g. motor vehicles, pedestrians). This type of traffic participant shall not be obstructed in its driving mission. Additionally, this type is allowed to move permanently within this area. The type of the reservation may be \textit{own-reserved} (= reserved for the type of traffic participant under consideration), \textit{externally-reserved} (= reserved for other type(s) of traffic participant) or \textit{equally-reserved} (= reserved equally between two different types of traffic participant). A link property identifies the area from which the reservation entitled traffic participants may come. Exceptions of these reservation types include restricted areas that are not reserved for any type of road user. In this case, the type of reservation is \textit{none}.

The \textit{overtake attribute} determines the permission to overtake other traffic participants within the given atomic behavior space.

As an outlook of this previous paper, we identified the necessity to represent interconnections between the individual atomic behavior spaces in order to describe the behavior relevant information of whole traffic networks. Within this contribution, we want to introduce the Behavior-Semantic Scenery Description (BSSD) that fulfills that and further requirements in order to represent the behavior space of whole traffic networks. Table \ref{tab:table1} summarizes all relevant terms as a basis for this work.

\begin{table}[!t]
\caption{Relevant Terms for this Work\label{tab:table1}}
\centering
\setlength{\tabcolsep}{3pt}
\begin{tabular}{| p{2cm} | p{4.9cm} | p{1cm} |}
\hline
\bf{Term} & \bf{Meaning} & \bf{Source}\\
\hline
Scenery & Geo-spatially stationary elements of the environment & \cite{Geyer.2014}\cite{Ulbrich.2015}\\
\hline
Regular Motion Space & Motion space for motor vehicles (usually the roadway) & \cite{Glatzki.91920219222021}\\
\hline
Behavior & Externally observable state changes & \cite{Nolte.2017}\cite{Czarnecki.2018b}\\
\hline
Behavior Space & Delimited set of legally possible behaviors & \cite{Glatzki.91920219222021}\\
\hline
Behavioral Demand & Limits of the behavior space & \ -\\
\hline
Behavioral Attribute & Dimension of the behavior space & \cite{Glatzki.91920219222021}\\
\hline
Atomic Behavior Space & Section of the scenery within which the behavioral attributes do not change & \cite{Glatzki.91920219222021}\\
\hline
\end{tabular}
\end{table}

\subsection{GOALS AND CHALLENGES}
From the basics of behavior spaces and behavioral attributes, it is evident that the behavior space represents the behavioral demands in semantic form. So far, by using only single, isolated behavior space the behavioral demands are only represented for sub-parts of the scenery without putting them into context with each other. However, the main goal of BSSD is to semantically represent the behavioral demands in the overall context of a considered scenery. Here, the behavioral demands apply to a specific type of traffic participant. If this main objective is achieved holistically, the following hypothesis may be corroborated and not be falsified \cite{Popper.1959}:

\textit{\textbf{Hypothesis}: The BSSD represents the behavioral demand of the scenery for a specific traffic participant in semantic form.}

For this work, the scope is the BSSD for an automated motor vehicle. Thus, given considerations and examples address HAV as a specific type of traffic participant. However, the BSSD can potentially be used for any type of traffic participant.
In the following, the sub-goals and challenges to achieve the stated main goal of this work are identified and discussed. Subsequently, these will be used as a basis for deriving the requirements for BSSD.

\textit{\textbf{Assignability}}: Currently, a description of individual atomic behavior spaces using behavioral attributes is possible based on a given scenery. At first, it is irrelevant whether the scenery is artificially generated or real. For the description of a behavior space, however, only the relevant scenery section is considered without establishing an explicit and traceable connection. For development, testing and operation of HAV it is necessary to know the connection of the behavioral demand to a real scenery or a real route network. In this way, for example, an ODD selection within a route network becomes possible. Thus, the goal is a traceable connection between real scenery and behavior space. This means that each (atomic) behavior space is assigned to its corresponding scenery section.

\textit{\textbf{Connectivity}}: In addition to unambiguously assigning behavior spaces to the scenery, it is necessary to establish the connection between the behavior spaces themselves. Initially, each behavior space exists independently of others. If an ODD of HAV is considered only within one atomic behavior space, information about a single atomic behavior space would be sufficient. Usually, the behavior space changes multiple times while moving through a road network due to changes in behavioral demands, for example, caused by traffic rules or various lane topologies. Thus, if an ODD contains multiple (different) atomic behavior spaces, the connection between them is essential. Even having only two different atomic behavior spaces requires an unambiguous connection, since both the entry into a new space and the associated driving in this space are linked to conditions. To fulfill these conditions, they must be known while being in the previous behavior space. Thus, the goal is a scenery description that enables the navigation through the individual atomic behavior spaces comparable to a map.

\textit{\textbf{Consistency}}: When assigning the behavior spaces and connecting them to each other, the absence of contradictions is another decisive factor. There must be no duplications or multiple references within the description. The description must provide contradiction-free and unambiguous behavioral information for each part of the scenery. This prevents parts of the scenery that should be described in the same way from a behavioral point of view from being represented differently in the description.

\textit{\textbf{Generality}}: Different use cases of HAV may require different ODD definitions and thus different associated sceneries to be navigated. To cover as many current and future use cases as possible, the BSSD should be generic. This means that an application is universally possible and in this way, every relevant scenery or ODD for the operation of automated vehicles can be mapped. Completeness is difficult to prove in this respect, but the goal should nevertheless be pursued with a view to the future of automated driving.

Based on the previously mentioned goals and the resulting challenges in developing the BSSD, requirements for the description are derived in the following.

\section{REQUIREMENTS FOR THE BSSD} \label{requirements}
First, the goal of assignability is considered. In order to unambiguously connect the scenery with the corresponding behavior spaces, the BSSD must first divide a scenery into individual parts that correspond to the atomic behavior spaces. An atomic behavior space usually corresponds to a lane segment, so the scenery must be broken down to the lane level. The first requirement (RQ) is therefore:

\textit{\textbf{RQ 1}: The BSSD shall divide the scenery into atomic behavior spaces.}

Once the scenery is divided into the individual parts corresponding to the atomic behavior spaces, the appropriate behavioral demands must be assigned. Thus, each individual part of the scenery shall have the four behavioral attributes allocated. The structure of an atomic behavior space as described in the basics has to be kept. Special attention has to be paid to the physical boundaries of the atomic behavior spaces, which have to be realized within the boundary attribute. These span the behavior space not only from a behavioral point of view but also from a geometric point of view. In summary the next requirement is:

\textit{\textbf{RQ 2}: The BSSD shall represent the associated behavioral attributes of the atomic behavior spaces.}

The goal of connectivity demands that not only individual atomic behavior spaces, but all behavioral demands in the entire road network are represented holistically. For this purpose, the atomic behavior spaces must be interconnected. It must be ensured that all behavioral demands of the individual atomic behavior spaces remain unchanged while establishing the connections. Consequently, no behavioral demands shall be added, nor may existing behavioral demands be removed or modified. As a result, there should be a navigable route network of atomic behavior spaces, so that the behavioral demands are explicitly given for each possible path within this network. Another constraint is the validity of the route network representation. The BSSD route network must represent the real route network, which is used to derive the BSSD, identically in the sense of navigability. This is the only way to enable later use of the BSSD for HAV development and operation (e.g. ODD specification or routing). Due to this endeavor, the following requirement is formulated:

\textit{\textbf{RQ 3}: The BSSD shall connect behavior spaces logically and consistently to a valid representation of the navigable route network.}

In order to achieve the goal of consistency, ambiguities must be excluded. Consequently, there must not be different descriptions for the same information content. It is possible that different scenery sections require the same behavior space, although they differ in the scenery characteristics. In these cases, the different scenery sections must each be assigned the same behavior space so that the information content is unambiguous and thus consistent. Neither assignability nor connectivity must suffer from the consideration of this condition. The following requirement is defined to fulfill the consistency:

\textit{\textbf{RQ 4}: If different sceneries impose the same behavioral demands, they shall always be represented by the same behavioral space.}

To meet the goal of generality, the BSSD should be as universally applicable as possible. This means that there should be no behavior space that cannot be represented by BSSD. Consequently, there must not be any real scenery or scenery section for which the behavior space cannot be represented or cannot be represented correctly. The final requirement is therefore:

\textit{\textbf{RQ 5}: The BSSD shall represent the behavior space to any real scenery.}

Taking into account the established goals, challenges and resulting requirements, the related work is identified and analyzed in the following.

\section{RELATED WORK}
It is apparent from the previous sections that two topics are relevant for the identification and analysis of related work: Scenery representation and behavior representation in the context of automated driving. Consequently, work is sought that addresses these topics and analyzed whether a link between both topics is established.

\subsection{SCENERY REPRESENTATION}
As a central component of the ODD, the scenery is used in the context of the representation of scenes, situations and scenarios. For the identification, derivation or generation of these representations, the description and representation of the scenery is indispensable.

A popular approach to represent sceneries is the application of ontologies, which are utilized to organize and structure knowledge. Bagschik \textit{et al.} \cite{Bagschik.2018b} employ this approach to generate and represent scenes for HAV by building an ontology based on a 5-layer model. They adapted the original 4-layer model for generic scenario description by Schuldt \cite{Schuldt.2013} and added another layer. The scenery is described using the four layers road-level, traffic-infrastructure, temporal changes of the previous layers and environment. Using the ontology, individual elements and their properties from these four levels are combined so that valid traffic sceneries are created. In the scene creation, the components such as individual lanes, hard shoulders or guard rails are represented as they are perceived in reality with no attached explicit rule or behavior information. These scene components are then arranged in a traffic-related manner. Traffic validity is thereby ensured with the knowledge base of state guidelines for highway construction within the ontology. In a further work, Bagschik \textit{et al.} \cite{Bagschik.2018} generate functional scenarios for the highway based on the same ontology. Both approaches are not applicable in this form for modeling urban sceneries or even intersections in general. Scholtes \textit{et al.} \cite{Scholtes.2021} adapt and extend the 5-layer model by the digital infrastructure layer for the structured description and classification of urban traffic and its surroundings, again not adding any explicit rule or behavioral information.

Ulbrich \textit{et al.} \cite{Ulbrich.2014} present a graph based information representation consisting of different hierarchical information layers. On the highest level, a topological representation of the road or intersection shows the different relationships and links between lanes and intersections. Lane boundaries link lane segments in lateral direction, while way points link lane segments in longitudinal direction. First explicit traffic rule information such as the permission to cross a boundary or speed limits of lane segments can be added. Still, information regarding behavioral rules are added without a systematic approach to cover every valid rule for the segments.

Buechel \textit{et al.} \cite{Buechel.2017} use a similar approach to describe road segments based on an ontology. A road segment consists of lane segments and lane markings arranged by statements such as \textit{isConnectedTo} or \textit{containsLeftLaneMarking}. Traffic regulations, e.g. country specific speed limits, are linked directly to road segments. In addition to modelling only road segments, Hülsen \textit{et al.} \cite{Hulsen.2011} build an ontology to describe traffic intersection situations. Similar to the approaches mentioned before, they use statements such as \textit{isRightOf} or \textit{hasRightOfWay} to describe the relationship between roads and crossings as well as between different traffic participants. Traffic signs and their meanings are included in the same manner. In another ontology-based approach, Regele \cite{Regele.2008} proposes a decision-making process of automated vehicles. By using a graph-like network of connected lanes, vehicles and objects, he considers explicit relations between lanes such as opposing traffic or bi-directional lanes by linking behavioral advises rather than explicit rules. For example, a behavioral advise for opposing traffic is take special care during crossing over the other lane.

Butz et al. \cite{Butz.2020} abstract traffic situations by using static zone graphs as an abstract scenery representation in order to perform a morphological behavior analysis. Zone graphs are constructed for one kind of scenery (e.g. a 4-way intersection or roundabout) and an HAV intention. They are made up of different types of zones connected by different types of edges. There exist driving zones for the HAV, position zones for other traffic participants or information zones that contain e.g. traffic signs. This approach is already strongly linked to a behavior representation. The behavioral part is not a pure representation of rules but already considers the abilities of the ego vehicle. By using a zwicky-box, equivalence classes for the required behavior are derived, e.g. a pedestrian crossing is either blocked or threatened (requirement is to stop in front). Within their representation not all traffic rules are covered, e.g. speed limits or overtaking bans are not considered.

Another popular approach to describe the scenery especially in simulation tools and operation are high definition maps. These maps represent the scenery in high detail with centimeter accuracy.  OpenDRIVE \cite{.2020} is a standardized file format developed for simulation applications, which require a precise description of road networks. This format structures roads, referred to a reference line with driving lanes and road features describing the scenery. Thus, it also includes traffic regulation elements such as traffic signs or lights. There also exists a traffic rule identifier, but without a unified method or structure to represent traffic or behavior rules at all.

A different approach is presented by Poggenhans \textit{et al.} \cite{Poggenhans.2018}. They extend and generalize the map format Liblanelet \cite{Bender.2014} to a new high definition map framework \textit{Lanelet2}. The basis of this format are lanelets, which are atomic road sections connected to each other forming the road network. Traffic rules as well as topological relationships do not change within a lanelet. Lanelet2 describes traffic rules using regulatory elements, which refer to elements that define these traffic rules (e.g. traffic signs or lights). Regulatory elements are referenced by at least one lanelet or area for which the linked traffic rules are valid. However, the regulatory elements do not represent the traffic rules holistically for the complete scenery. For example, there is no information about lanes that are driven against their intended direction. Nevertheless, this information is needed for overtaking, among other things. Furthermore, the regulatory elements do not represent the traffic rules in a uniform way. All traffic rules, regardless of type, are represented using this one class.

\subsection{BEHAVIOR REPRESENTATION}
The behavior of an automated vehicle must not only be traffic rule compliant (locally as well as globally), but is also constraint by other aspects, e.g. local culture or comfort constraints. These constraints may be in conflict with each other. Censi et al. \cite{Censi.2019} present a behavior specification method with so-called \textit{rulebooks}. Not all rules can be satisfied simultaneously. With the help of the rulebooks the different rules are arranged in a hierarchy (rulebook = preordered set of rules). A rule is a scoring function (possibility of more severe violation than other violation) that helps to specify and order behavior constraints. By giving safety the highest weighting, country and culture specific rules can then be considered at a lower level. However, the work does not provide representations of the individual rules, e.g. traffic rules.

For the representation of individual rules, on the one hand based on the surrounding scenery, it is possible to derive the present traffic rules and with that the required behavior of HAV. But on the other hand, there are different approaches to represent the required behavior on a global level without considering the present scenery. Most prominently, Shalev-Shwartz \textit{et al.} \cite{ShalevShwartz.8212017} introduce the Responsibility-Sensitive Safety (RSS) model. In this approach, five “common sense” driving rules are formalized in a mathematical model. The authors prove that if every traffic participant follows these mathematically formalized rules no collisions would occur. While considering basic priority rules at intersections, most of the traffic regulations based on local scenery (e.g. traffic lights, speed limits) are not represented in this approach. Further approaches to formalize global traffic rules are presented by Rizaldi \cite{Rizaldi.2017} and Esterle \cite{Esterle.18.11.202016.12.2020}.

Stolte \textit{et al.} \cite{Stolte.2017} perform a hazard analysis and risk assessment for an unmanned protective vehicle which results in safety goals for the development of HAV. These goals state behavioral requirements, also regarding traffic rules (e.g. “overrunning hard shoulder markings must be prevented”), but remain unconnected to a representation of the scenery.

The National Highway Traffic Safety Administration (NHTSA) \cite{Thorn.2018} and Waymo \cite{Waymo.2020} present behavioral competencies recommended for the development and testing of HAV, but these competencies are to abstract to serve as an explicit behavioral demand representation in a route network.

Lopez \textit{et al.} \cite{PerdomoLopez.2017} use flow charts to model the required behavior while driving through urban intersections resulting in nine decision points.  These flow charts represent various queries regarding the present scenery and planned maneuvers and based on this give out the present priority rules. The charts are not linked to a specific intersection but rather need to be reapplied for every intersection to receive the resulting rules. 

Behavior planers implemented in HAV need to convert the behavior rules of the surrounding scenery into a valid behavior conforming to these rules. Therefore, these planers need to represent this behavioral demand in order to work with it. To the knowledge of the authors no unified approach to do this extists. Pek \textit{et al.} \cite{Pek.2020} introduce an online verification method to reduce accidents caused by automated vehicles. They consider every legally possible behavior of traffic participants considering the dynamic feasibility. In order to allow an online verification, the considered traffic rules were formalized in a previous step. How these traffic rules are derived and represented remains unclear. All these behaviors are collected in one behavior set which corresponds to one specific traffic participant.

\subsection{SUMMARY}
The related work shows various approaches to model the surrounding scenery. With current methods the elements as perceived by the human eye are represented in ontologies to model the relations in between them. This representation of the element rather than the information that the element carries results in a big amount of data that needs to be interpreted in order to use it for the automated driving task. Lanelet2 \cite{Poggenhans.2018} seems to be a promising approach to explicitly represent traffic rules, but still lacks to represent the explicit demanded behavior for a scenery. So far, the derivation and representation of the behavior of HAV is based on traffic rules, common sense rules or safety analyses. Expert knowledge is often used in addition to complement the resulting behavior specification.

Overall, scenery and behavior are often represented in separate ways, which usually requires at least one additional step in the derivation process. There is no approach specifying the demanded behavior based on the scenery within a HAV has to operate. Furthermore, the different dimensions of behavior rules are not examined or modeled separately. Therefore, the requirements derived in Section \ref{requirements} are only partly and not holistically met in the current related work. As a result, we investigate a new semantic description, directly linking the scenery with its demanded driving behavior.

\begin{figure*}[!b]
\centering
\includegraphics[width=\textwidth]{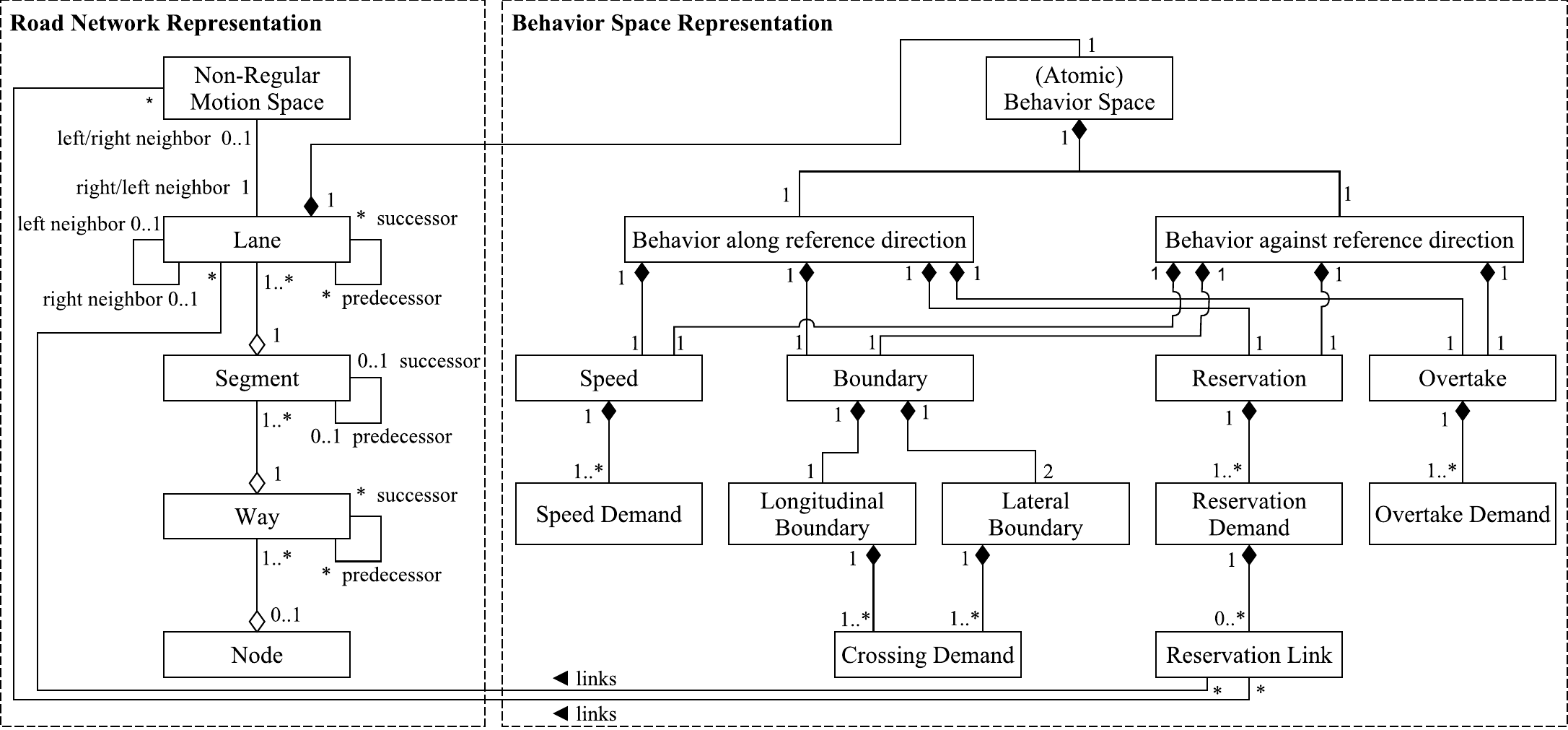}
\caption{Structure of the Behavior-Semantic Scenery Description.}
\label{fig_1}
\end{figure*}

\section{BEHAVIOR SEMANTIC SCENERY DESCRIPTION} \label{bssd}
In the following, the elements necessary for a BSSD and their relationship to each other are derived from the identified requirements. With regard to an implementation, the structure of the BSSD should be as generic as possible and thus independent of the target format or target system. This ensures that the BSSD is utilizable in any use case and ODD. The aim is to achieve a description that represents all necessary elements and properties of the BSSD such that the requirements from Section \ref{requirements} are met. Fig. \ref{fig_1} represents the generic structure of the BSSD resulting from the derived necessary elements and their relationship to each other.

\subsection{ELEMENTS FOR THE ROAD NETWORK REPRESENTATION}
It follows directly from RQ 1 that the basis for a BSSD is a (partial) route network that is decomposed according to atomic behavior spaces. In lateral extension, a lane represents the smallest possible road space onto which an atomic behavior space is represented. For this purpose, it must necessarily be possible to represent individual lanes. In addition to a conventional lane for motor vehicles, a bicycle lane, for example, may also represent a lane. Such a lane is potentially used by a motor vehicle as well. Besides lanes within the regular motion space, elements of non-regular motion space have to be considered for the representation of reservation links (e.g. pedestrians coming from a sidewalk onto a pedestrian crossing).

Depending on the use case, it may not be sufficient to represent individual atomic behavior spaces in isolation. They must be considered in the overall context of a road network so that RQ 3 is satisfied to ensure connectivity. In terms of navigability, all possible driving options as they exist in reality must therefore be represented. For every point in the route network where multiple driving options follow, the available behavior spaces must be represented. Since geometry is not a part of the description of a behavior space, the BSSD in its plain form does not require any geometry for the representation of sceneries. In this case, further auxiliary elements besides lanes are necessary for a consistent route network representation. If the BSSD is integrated into a map containing geometric information, some of these auxiliary elements may be omitted, depending on the level of detail of the map. For example, the relationship of individual lane sections in a HD map would be evident based on geometric adjacency alone, without the need to define further dependencies. Since a representation entirely without geometry requires the most auxiliary elements, this case is considered below. If geometric information is added, the corresponding auxiliary elements can simply be neglected. If they are beneficial for the application, however, it is still possible to use them.

Route networks can be described without geometry by a logically constructed topology following the topological graph theory. A road network is represented, as is common in navigation, using nodes and edges. The \textit{nodes} represent traffic points where the traffic flow branches in different directions. In the scenery, these points correspond to intersections, traffic circles or junctions, for example. All connecting roads between the nodes are modeled as edges, which are called \textit{ways} in the following. Consequently, more than two ways are connected at nodes. Within nodes, again ways represent the possible connections between the incoming and outgoing ways adjacent to the nodes. Each way in a road network therefore may have arbitrarily many predecessors or successors. This ambiguity of nodes is explicitly desired, because in this way the different driving options at nodes are represented. However, for a lane-accurate representation of the scenery, the ways must be further subdivided into \textit{lanes}. As soon as different lane topologies prevail within a way (e.g. transition to a different number of lanes), a subdivision of the lanes in longitudinal direction becomes necessary.

For lateral transitions between lanes (e.g. lane changes) the neighbors of a lane are specified. In order to ensure uniqueness in lateral transitions every lane has only one left and right neighbor at most. This results in a longitudinal segmentation of a way into a \textit{segment} whenever any lane has a change in its behavior space. In order to enable the linkage of reservation receiving traffic participants lanes may have \textit{non-regular motion space} as a left or right neighbor. In contrast to lateral neighbors, a lane may have any number of predecessors or successors in longitudinal direction. As with ways at nodes, this property allows the assignment of multiple driving options for diverging or separating lanes and the associated atomic behavior spaces.

An advantage of segmentation is the holistic representation of behavioral demands within a road segment. A segment represents the behavior space across the entire lane width. In this way, all behavioral demands for driving on the road section are explicitly available. The same principle applies to a way, which in turn consists of at least one segment.

In summary, depending on the integration of geometric information, the elements listed in Table \ref{tab:table2} are necessary for mapping the BSSD to a road network. The resulting structure of the road network representation within the BSSD is shown in Fig. 1 on the left-hand side.

\begin{table}[!t]
\caption{Necessary Elements for BSSD of a Road Network\label{tab:table2}}
\centering
\setlength{\tabcolsep}{3pt}
\begin{tabular}{ | p{1.7cm} | p{6.4cm}|}
\hline
\bf{Term} & \bf{Description}\\
\hline
Node & Area in which multiple ways interfere and incoming and outgoing ways are connected.\\
\hline
Way	& Connecting road between and within nodes.\\
\hline
Segment & Section of a way in which the mapped behavior space is constant in longitudinal direction.\\
\hline
Lane & Section of a segment in which the mapped behavior space is constant (no change at all).\\
\hline
Non-regular motion space & Area outside of the regular motion space.\\
\hline
\end{tabular}
\end{table}

\subsection{ELEMENTS FOR THE BEHAVIOR SPACE REPRESENTATION}
After the atomic behavior spaces can be represented using the elaborated structure for a valid representation of road networks (RQ 1 and RQ 3), a structure for mapping the behavioral demands onto the atomic behavior spaces has to be derived (RQ 2). This structure must additionally fulfill RQ 4 to achieve consistency.

Due to the directionality of the behavioral demands, the atomic behavior space must always be able to represent both possible driving directions of an HAV. Therefore, an \textit{atomic behavior space} always consists of two additional elements, the \textit{behavior along reference direction} and the \textit{behavior against reference direction} (the reference direction may be selected as desired). Both directions must cover the same knowledge requirements about the possible behavioral demands: \textit{What is the speed limit? What conditions apply when changing lanes or entering a new space? Which road users must be given priority? Is overtaking allowed?}

As a result, for both considered driving directions, the behavioral attributes \textit{speed}, \textit{boundary}, \textit{reservation} and \textit{overtake} are each assigned exactly once. In turn, the behavioral attributes always belong to only one considered driving direction within an atomic behavior space. The behavioral demands describe the characteristic of the individual behavioral attributes in order to fulfill the mentioned knowledge requirements. They are stored as a part of the respective attribute.

\textit{Speed Attribute}: At least one \textit{speed demand} element must be defined, specifying the maximum allowed driving speed within the atomic behavior space. Additional demand elements may be defined for speed limits under certain conditions such as time of day or weather. A required minimum speed may be added as well.

\textit{Boundary Attribute}: The behavioral demands are restricted to crossing conditions of the respective boundaries. An atomic behavior space always consists of one \textit{longitudinal (entry) boundary} and two \textit{lateral (exit) boundaries}. At least one or more \textit{crossing demand} elements are assigned to each of the three boundaries. Conversely, each crossing demand element is part of a boundary element. An example for a double assignment of a longitudinal boundary is a stop line at a traffic light system. Here, different crossing demands apply for active or inactive traffic lights.

\textit{Reservation Attribute}: As introduced in Section \ref{preliminaries}, the reservation attribute covers all behavioral demands regarding priority and residence allowance rules. By abstracting the description of these demands, it is possible to apply the representation to all atomic behavior spaces independent of the type of road section (e.g. junction, road, roundabout) that is described. At least one \textit{reservation demand} element is assigned to the reservation attribute. Dependent on the type of reservation (own, externally, equally, none) further elements are required. For the externally- and equally-reserved cases, the type of the reservation-entitled road users must be represented. Additionally, there is the \textit{reservation link} element, which indicates the origin and, if necessary, the destination direction of these road users by directly referring the respective lane element. Any number of reservation links can be defined for the reservation demand, which can address any number of lane or non-regular motion space elements.

\textit{Overtake Attribute}: The overtake attribute has at least one \textit{overtake demand} element. As with the speed attribute, an overtake prohibition may be linked to different conditions, resulting in multiple overtake demands.

The resulting structure of the behavior space representation within the BSSD is shown in Fig. \ref{fig_1} on the right hand side. With the elaborated structure it is possible to assign a complete behavior space to each scenery section (RQ 2). The basis for the interconnection of the individual atomic behavior spaces (RQ 3) is the structure of the road network derived in the previous section. The resulting overall structure (Fig. \ref{fig_1}) now represents not only each individual behavior space, but also their concatenations. Thus, the behavioral demands resulting from subsequent behavior spaces are represented and their sequence is directly accessible.

\section{APPLICATION AND REAL WORLD EXAMPLES}
In this section, the generic BSSD from the previous section is instantiated to describe real world sceneries. For this purpose, the BSSD is created using the map framework Lanelet2 \cite{Poggenhans.2018} as a basis. Two real scenery sections in Darmstadt (Germany) are thereby considered and explained in the following. These examples demonstrate that despite striking differences in the two scenery sections, the resulting BSSD shows only few differences. The two examples additionally serve as an evaluation of the approach by checking the requirements specified in Section \ref{requirements}. The visualization of the scenery sections was done with JOSM \cite{JOSM.2021} using the Lanelet2 map style. In addition, some of the BSSD information is visualized using the BSSD map style. Information boxes, arrows and pictograms were added manually to visually represent further information stored in the BSSD map implementation. The blue and black circles with respective numbering visually support the explanations in the following.

Before describing the examples themselves, the structure of the Lanelet2 framework and its impact on the BSSD implementation is first explained for further understanding. Lanelet2 builds on the map format OpenStreetMap (OSM) \cite{Haklay.2008}, which uses the elements \textit{node}, \textit{way} and \textit{relation} in order to model a map (these nodes and ways are different to the defined ones in Section \ref{bssd}, Table \ref{tab:table2}). Ways consist of nodes and correspond to \textit{linestrings} in Lanelet2. Relations refer to \textit{members} like linestrings, nodes or relations and assign a \textit{role}. The role defines the property or relationship of the member with respect to the relation.

Lanelet2 maps are augmented with BSSD information for the application of BSSD, while fully preserving the functionality of the original map. The core element of the Lanelet2 map format are \textit{lanelets}, which are used as atomic components of road networks to build maps. They are modeled as relations and always reference two lateral boundaries in the form of linestrings with the roles \textit{left} and \textit{right}, within which directed movements take place and traffic rules do not change. Thus, for a motor vehicle, lanelets generally represent a lane section of a roadway. The representation of bicycle lanes or crosswalks as well as non-regular motion space is additionally possible.

The construction of a lane network for the BSSD is not necessary when using Lanelet2, since the map already provides the necessary information. Nevertheless, it must be ensured that the assignment of atomic behavior spaces, as shown in the generic UML representation of the BSSD (Fig. \ref{fig_1}), to this route network is possible. In case two different behavior spaces have to be assigned to a single lanelet, this lanelet can be split considering the design rules of Lanelet2. If the Lanelet2 map is to remain untouched, a lanelet can be artificially split using additional BSSD elements in OSM format. If an atomic behavior space contains two or more lanelets, they are referenced together without changing the format itself. The union of multiple lanelets would break the Lanelet2 format and make it unusable at this point. To represent the behavioral demands of the longitudinal boundary, most lanelets also require additional linestrings. However, these can be added in a Lanelet2-compliant way without endangering the format. If such linestrings are already available, e.g. in the form of stop lines at the correct position, no new elements have to be created.

\begin{figure*}[!t]
\centering
\includegraphics[width=\textwidth]{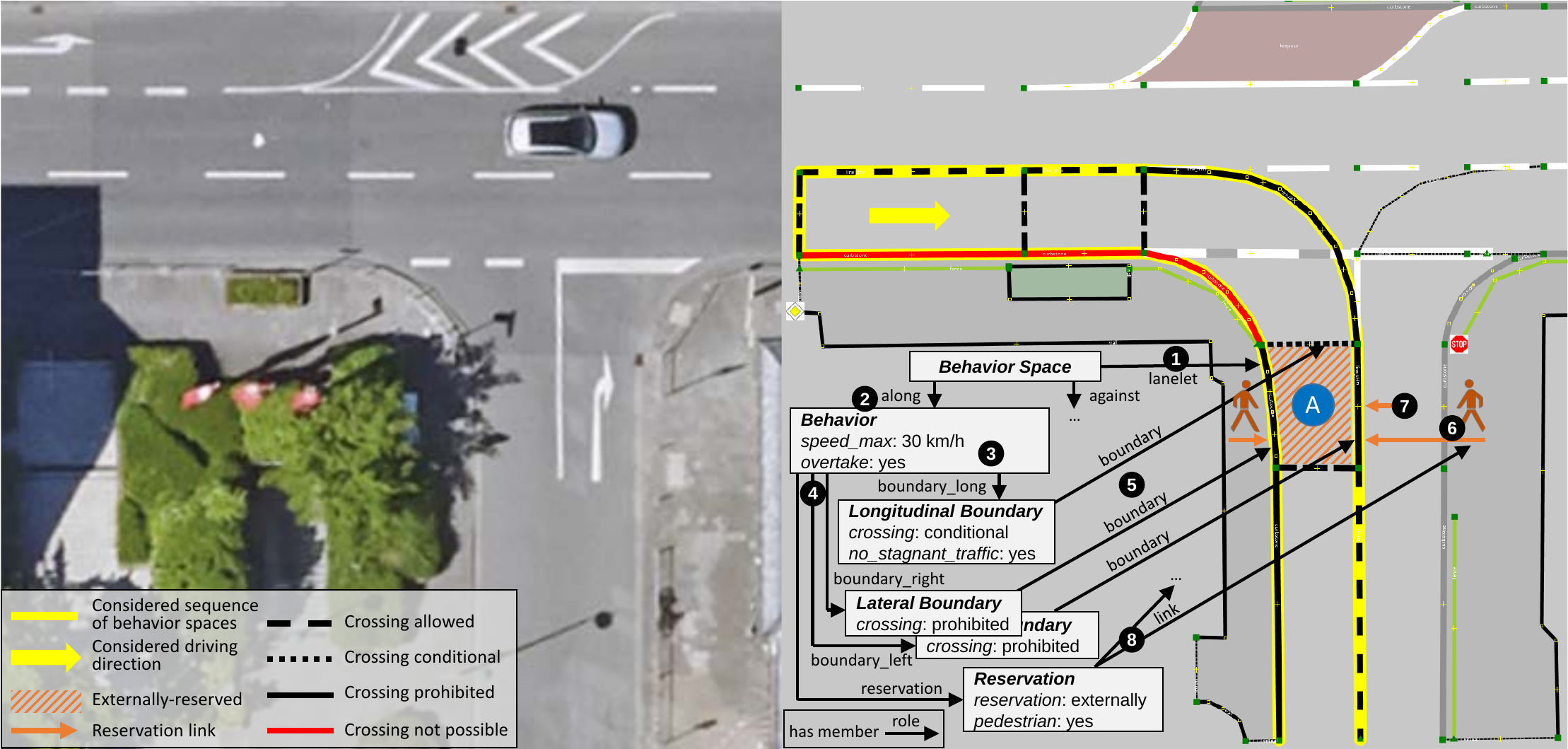}
\caption{Example A: T-junction in Darmstadt, Germany (Aerial image $\copyright$~Orthophoto Vermessungsamt Darmstadt 2021).}
\label{fig_2}
\end{figure*}

After introducing the basics of the Lanelet2 framework and its effects on the BSSD implementation, more detailed explanations regarding the BSSD implementation are provided considering concrete examples in the following. Fig. \ref{fig_2} shows the aerial image and the corresponding Lanelet2 map with the BSSD extension of example A. This example represents a T-junction within a 30 km/h speed zone. The priority road is a two lane one-way road and the secondary road is a two-lane road with bidirectional traffic. In order to explain the implemented structure, only one sequence of atomic behavior spaces is considered (yellow marking), leaving the other behavior spaces unrepresented. Following the sequence in the marked direction (yellow arrow) is equivalent to a right turn maneuver. We consider the atomic behavior space A (blue circle) for a detailed explanation of the BSSD information. It must be noted that pedestrians that potentially cross the secondary road during that right turn maneuver only can cross the road in the considered behavior space. A fence (light green line) prevents the crossing in the preceding behavior spaces. A crossing in the successive behavior space would no longer be part of the turn maneuver resulting in different behavioral demands. Of course, the global behavioral rules require that even collisions with pedestrians climbing over the fence are avoided. Again, at this point it should be noted that the focus of this work is on local behavioral rules that arise from specific sceneries.

Atomic behavior spaces are directly mapped to their corresponding lanelet (black circle 1), as the behavioral demands change before and after this lanelet. The lanelet is defined as a member with the role \textit{lanelet} of this \textit{behavior space}, so that the scenery linkage is directly established. As further member, the behavior space has the relation \textit{behavior} with the role \textit{along} (black circle 2), which represents accordingly the behavior along the reference direction (the reference direction is defined by the lanelet). Besides the type of the relation, which is always defined, the behavioral demands of the attributes speed and overtake are directly stored within this relation (black circle 3). For behavior space A, the maximum allowed speed is 30 km/h and overtaking is not prohibited.

The behavioral demands of the remaining behavioral attributes boundary and reservation are modeled as relations. They are members of behavior with the respective role (\textit{boundary\_long}, \textit{boundary\_right}, \textit{boundary\_left} and \textit{reservation}) as highlighted by the black circle 4. These elements in turn reference the Lanelet2 map information. For example, the boundaries are directly linked to the linestrings of the lanelets or the newly created linestrings for the longitudinal boundaries (black circle 5). Likewise, the linking of lanelets, from which road users with reservation claims may come, takes place. In this example, when entering the considered behavior space, priority must be given to pedestrians coming from the sidewalks to the left and right (black circle 6). That behavioral demand is a result of the turn maneuver since motor vehicles and bicycles generally have to give priority to pedestrians crossing the street while turning. Since crossing pedestrians might already be on the road in the lateral adjacent lanelet, this area has to be considered as a link as well (black circle 7). In general, all areas that have to be crossed by reservation entitled traffic participants must be considered and linked to the according reservation element. Thus, the reservation demand of the considered behavior space is \textit{externally-reserved} for pedestrians with a reservation \textit{link} to the corresponding Lanelet2 elements as highlighted by black circle 8 (sidewalks and adjacent lanelet, indicated by the orange arrows and pictograms). Only one of the three links is presented explicitly in this example for reasons of clarity.

\begin{figure}[!t]
\centering
\includegraphics[width=\linewidth]{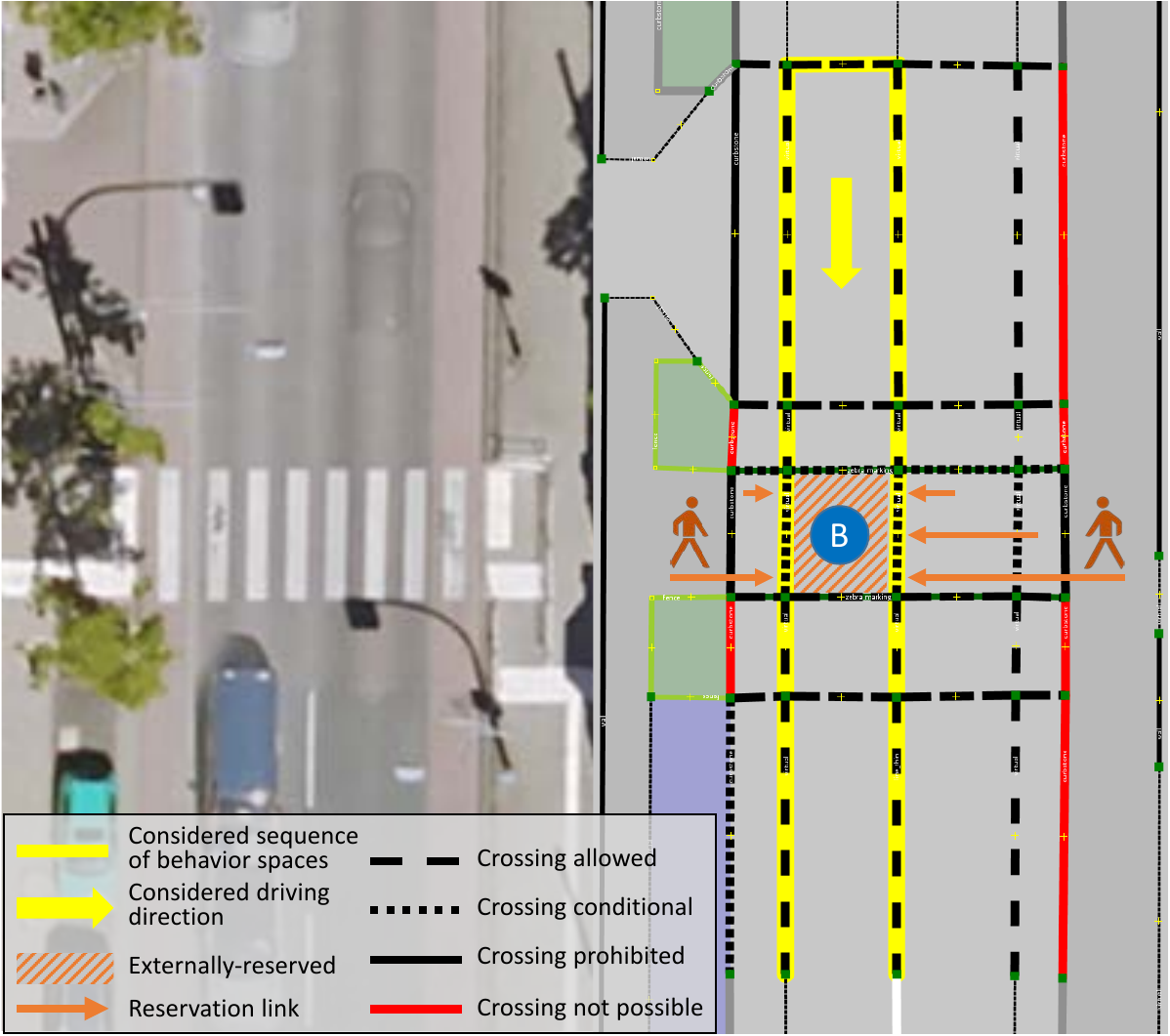}
\caption{Example B: Two-lane road with crosswalk in Darmstadt, Germany (Aerial image $\copyright$~Orthophoto Vermessungsamt Darmstadt 2021).}
\label{fig_3}
\end{figure}

A second real scenery section is considered in example B in Fig. \ref{fig_3} that shows a two-lane road with bidirectional traffic in a 50 km/h speed zone, lateral adjacent bicycle protection lanes and a crosswalk. A parking area adjacent to one bicycle protection lane is found as well (blue colored area in the Lanelet2 map). In this example, the behavioral demands regarding the boundary elements of the behavior spaces are visualized considering a driving direction as indicated by the yellow arrow. We again consider a certain sequence of atomic behavior spaces (yellow marking) and, in particular, the atomic behavior space B representing a lane section on the crosswalk.

In order to demonstrate the benefits of the BSSD due to an usage of only four behavioral attributes to describe the behavioral demands, we directly compare the presented atomic behavior spaces A and B. The corresponding behavioral demands are shown in Table \ref{tab:table3} and can be analyzed attribute-wise.

We want to focus on the behavioral demands regarding longitudinal boundary and reservation since lateral boundaries are less relevant considering the yellow marked sequence. In both examples the crossing demand of the longitudinal boundary is \textit{no stagnant traffic}. This means that entering the behavior space is only allowed if the space can be passed without coming to a rest, no matter where this rule arises from. In example A that demand results from being part of a turn maneuver at an intersection. The same demand in example B results from being part of a crosswalk. Thus, a HAV would have to check if sufficient driving space is available in the successive atomic behavior spaces before entering the considered space. Additionally, both examples have the same reservation demands. In both cases, HAV entering this behavior space have to give priority to pedestrians coming from the sidewalks and areas that they have to pass by crossing the road as well. In both examples, HAV must check the laterally adjacent areas for pedestrians potentially crossing the street. Furthermore, staying in the behavior spaces A and B is prohibited, meaning that this area must be left as soon as possible. This behavioral demand is also indicated by the external reservation. The attributes of speed and overtaking of example A and B differ in their behavioral demands as well as the crossing demands of the lateral boundaries.

\begin{table}[!t]
\caption{Behavioral Demands of Example A and B\label{tab:table3}}
\centering
\setlength{\tabcolsep}{3pt}
\begin{tabular}{| p{1.5cm} | p{3.2cm} | p{3.2cm} |}
\hline
\bf{Attribute} & \bf{Behavior Space A} & \bf{Behavior Space B}\\
\hline
Speed &  
\textit{max}: 30 km/h & 
\textit{max}: 50 km/h\\
\hline
Boundary & 
\textit{long.}: conditional \newline 
\hangindent=.3in no stagnant traffic, &
\textit{long.}: conditional \newline 
\hangindent=.3in no stagnant traffic, \\
& \textit{left}: prohibited, &
\textit{left}: conditional \newline 
\hangindent=.23in no stagnant traffic, \\
& \textit{right}: prohibited & 
\textit{right}: conditional \newline
\hangindent=.28in no stagnant traffic\\
\hline
Reservation & 
\textit{type}: externally-reserved, \newline
\textit{object}: pedestrians & 
\textit{type}: externally-reserved, \newline
\textit{object}: pedestrians\\
\hline
Overtake & 
\textit{permission}: yes & 
\textit{permission}: no\\
\hline
\end{tabular}
\end{table}

However, the equality of the behavioral demands of longitudinal boundary and reservation show that completely different sceneries may indeed be very similar in their demands. These similarities potentially result in reduced development and testing effort of HAV, since the diversity at the behavioral level is reduced. A behavior planner would need to perform some of the same automated driving tasks in the examples shown, which would not have been apparent based on the scenery itself. Resulting from this, the test criteria for tests of the behavior planer would be the same for the equivalent demands on both sceneries and thus, reusable in between them in order to save effort. It would be even possible to perform an online check of compliance to this criteria during operation when localizing within the BSSD map. Additionally, capability-based route planning can consider the current skills of the HAV with regard to the behavioral demands. In the first example, a route planer could decide to not turn right, if the vehicle does not have the skill to yield for pedestrians (no matter if this is principally the case or a function of the vehicle is degraded).

In both examples, concrete (navigable) sequences of atomic behavior spaces were considered. Nevertheless, for both scenery sections all behavior spaces are available in the sense of a complete BSSD (we will provide the complete BSSD of the example as open source). Thus, the decomposition of the scenery into atomic behavior spaces (RQ 1), the assignment of the corresponding behavioral attributes (RQ 2) and also the connection of the individual behavior spaces are addressed (RQ 3). The comparison of the two examples shows that behavior spaces of the BSSD can be very similar despite strongly different sceneries. The reservation dimension of both behavior spaces is identical. This shows, that even with completely different sceneries, RQ 4 is addressed, so that different sceneries with the same behavioral demands are  represented in the same way. So far, we have transferred scenery sections from Darmstadt and Karlsruhe into a BSSD using our Lanelet2 extension without any complications. Additionally, we have a BSSD instantiation in topological form (without geometry information) by using an ontology. We will also provide these other examples open source. While these examples do not prove that BSSD works for every conceivable (HAV-relevant) scenery (RQ 5), they do not falsify the application of BSSD. Thus, the hypothesis stated in Section \ref{requirements} is corroborated based on current evidence.

\section{CONCLUSION AND OUTLOOK}
In this work, we introduced a generic concept for a Behavior-Semantic Scenery Description (BSSD) for automated driving. By applying an instance of this concept to real word examples, we demonstrated that it is applicable to real world sceneries. So far, there was no approach in scientific publications that holistically assigned the behavioral demand of an automated vehicle to the scenery.

This representation of the behavioral demand is a highly important information that current scenery descriptions and representations where lacking to provide. With our approach, we explicitly provide the scenery-based information that is relevant to the driving task of automated vehicles. This includes all relevant relations between different sections of the road network. As a result, the BSSD represents what local behavioral constraints need to be fulfilled and where within a present scenery this needs to be accomplished.

The BSSD is structured in an abstract and unified way, such that we see the possibility that the description will be applicable to any traffic area relevant for the operation of automated vehicles. This universality needs to be researched and confirmed in future investigations.

A large field of possible use cases and applications for the usage of the BSSD opens up. BSSD paves the way for safety by design enabling the development-relevant derivation of requirements for the vehicle's driving behavior directly from a defined or yet-to-be-defined ODD. In addition, due to the unambiguous scenery linkage of these requirements, a possibility for stepwise testing and validation of sub-areas of the ODD would be created, as it is aimed for in the UNICAR\textit{agil} project \cite{Woopen.2018}. With respect to testing, a verification of traffic rule compliance of HAV becomes possible.

In addition to these advantages in development and testing of HAV, other potential applications are arising for driving operations. Both route planning and trajectory planning could benefit from the explicit knowledge of behavioral demands in road traffic. Routes could be planned in an ODD-compliant manner so that no route section is driven for which the necessary driving capabilities are not available. Behavior planners would have more explicit input and would not have to interpret and derive the behavior rules themselves using conventional input data.

We will make the format and different instances of the BSSD available on GitLab\footnote{https://gitlab.com/tuda-fzd/scenery-representations-and-maps/behavior-semantic-scenery-description} and GitHub\footnote{https://github.com/tuda-fzd/behavior-semantic-scenery-description}. The presented examples will be published as well.

\section*{ACKNOWLEDGMENT}
M. L. especially thanks the colleagues at the Technical University of Darmstadt and Technical University of Braunschweig for their great coorporation and support within UNICAR\textit{agil}.

F. G. kindly thanks Continental for their great cooperation and support within PRORETA 5, a joint research project of the University of Bremen, Technical University of Darmstadt, TU Iasi and Continental to investigate future concepts for autonomous driving systems.

We want to particularly thank Eric Krämer for his active support during the implementation of BSSD.

\bibliographystyle{IEEEtran}
\bibliography{bibliography}

\begin{IEEEbiography}
{Moritz Lippert} finished his Bachelor of Science and Master of Science Degrees in Mechanical and Process Engineering at Technische Universität Darmstadt. Since 2018 he is a Research Associate at the Institute of Automotive Engineering at Technische Universität Darmstadt where he is currently pursuing his PhD. In his main research topic, the safety of automated driving, he investigates the derivation of scenery-based requirements for the automated driving task and capability-based route planning.
\end{IEEEbiography}

\begin{IEEEbiography}
{Felix Glatzki} finished his Bachelor of Science and Master of Science Degrees in Mechanical and Process Engineering at Technische Universität Darmstadt. Since 2019 he is a Research Associate at the Institute of Automotive Engineering at Technische Universität Darmstadt where he is currently pursuing his PhD. In his main research topic, the safety of automated driving, he investigates the description and verification of traffic rule compliance for automated vehicles.
\end{IEEEbiography}

\begin{IEEEbiography}
{Hermann Winner} began working at Robert Bosch GmbH in 1987, after receiving his PhD in physics, focusing on the predevelopment of “by-wire” technology and Adaptive Cruise Control (ACC). Beginning in 1995, he led the series development of ACC up to the start of production. Since 2002 he has been pursuing the research of systems engineering topics for driver assistance systems and automated driving as Professor of Automotive Engineering at the Technische Universität Darmstadt. He discovered the “approval trap” of autonomous driving, the still unsolved challenge to validate safety of autonomous driving before market introduction.
\end{IEEEbiography}

\EOD

\end{document}

%% file: copyright.tex
\thispagestyle{empty}
\pagestyle{empty}
\twocolumn[
\begin{@twocolumnfalse}
	
	{\Large This work has been submitted to the IEEE for possible publication in \emph{IEEE Access}.} \\ \\

	Cite as:
	\vspace{0.1cm}
	
	\noindent\fbox{%
		\parbox{\textwidth}{%
			M.~Lippert, F.~Glatzki, and H.~Winner, ``{Behavior}-{Semantic} {Scenery} {Description} ({BSSD}) of {Road} {Networks} for {Automated} {Driving},'' {submitted for publication}. 
			
		}%
	}
	\vspace{2cm}
	
\end{@twocolumnfalse}
]

\noindent\begin{minipage}{\textwidth}
	
	\hologo{BibTeX}:
	\footnotesize
	\begin{lstlisting}[frame=single, breakatwhitespace=true]
		@article{lippert_behavior_2022,
			author={{Lippert}, Moritz and {Glatzki}, Felix and {Winner}, Hermann},
			title={{Behavior}-{Semantic} {Scenery} {Description} ({BSSD}) of {Road} {Networks} for {Automated} {Driving}},
			year={2022},
			publisher={submitted for publication}
		}
	\end{lstlisting}
\end{minipage}

%% file: main.bbl
\begin{thebibliography}{10}
\providecommand{\url}[1]{#1}
\csname url@samestyle\endcsname
\providecommand{\newblock}{\relax}
\providecommand{\bibinfo}[2]{#2}
\providecommand{\BIBentrySTDinterwordspacing}{\spaceskip=0pt\relax}
\providecommand{\BIBentryALTinterwordstretchfactor}{4}
\providecommand{\BIBentryALTinterwordspacing}{\spaceskip=\fontdimen2\font plus
\BIBentryALTinterwordstretchfactor\fontdimen3\font minus
  \fontdimen4\font\relax}
\providecommand{\BIBforeignlanguage}[2]{{%
\expandafter\ifx\csname l@#1\endcsname\relax
\typeout{** WARNING: IEEEtran.bst: No hyphenation pattern has been}%
\typeout{** loaded for the language `#1'. Using the pattern for}%
\typeout{** the default language instead.}%
\else
\language=\csname l@#1\endcsname
\fi
#2}}
\providecommand{\BIBdecl}{\relax}
\BIBdecl

\bibitem{ISO.2020}
{International Organization for Standardization}, ``{ISO/TR 4804: Road vehicles
  - Safety and cybersecurity for automated driving system - Design,
  verification and validation}.''

\bibitem{SAEInternational.2014}
{SAE International}, ``{Taxonomy and Definitions for Terms Related to On-Road
  Motor Vehicle Automated Driving Systems},'' 2014.

\bibitem{BundesministeriumderJustizundfurdenVerbraucherschutz.2013}
{Bundesministerium der Justiz und f{\"u}r den Verbraucherschutz},
  ``{Stra{\ss}enverkehrs-Ordnung: StVO},'' 2013.

\bibitem{ShalevShwartz.8212017}
\BIBentryALTinterwordspacing
S.~Shalev-Shwartz, S.~Shammah, and A.~Shashua, ``On a formal model of safe and
  scalable self-driving cars.'' [Online]. Available:
  \url{http://arxiv.org/pdf/1708.06374v6}
\BIBentrySTDinterwordspacing

\bibitem{SOTIF}
{International Organization for Standardization}, ``{ISO 21448:2022 - Road
  vehicles - Safety of the intended functionality},'' 2022.

\bibitem{Poggenhans.2018}
F.~Poggenhans, J.-H. Pauls, J.~Janosovits, S.~Orf, M.~Naumann, F.~Kuhnt, and
  M.~Mayr, ``Lanelet2: A high-definition map framework for the future of
  automated driving,'' in \emph{2018 IEEE Intelligent Transportation Systems
  Conference}.\hskip 1em plus 0.5em minus 0.4em\relax Piscataway, NJ: IEEE,
  2018, pp. 1672--1679.

\bibitem{Geyer.2014}
S.~Geyer, M.~Baltzer, B.~Franz, S.~Hakuli, M.~Kauer, M.~Kienle, S.~Meier,
  T.~Wei{\ss}gerber, K.~Bengler, R.~Bruder, F.~Flemisch, and H.~Winner,
  ``Concept and development of a unified ontology for generating test and
  use--case catalogues for assisted and automated vehicle guidance,'' \emph{IET
  Intelligent Transport Systems}, vol.~8, no.~3, pp. 183--189, 2014.

\bibitem{Ulbrich.2015}
S.~Ulbrich, T.~Menzel, A.~Reschka, F.~Schuldt, and M.~Maurer, ``Defining and
  substantiating the terms scene, situation, and scenario for automated
  driving,'' in \emph{2015 IEEE 18th International Conference on Intelligent
  Transportation Systems (ITSC 2015)}.\hskip 1em plus 0.5em minus 0.4em\relax
  Piscataway, NJ: IEEE, 2015, pp. 982--988.

\bibitem{Nolte.2017}
M.~Nolte, G.~Bagschik, I.~Jatzkowski, T.~Stolte, A.~Reschka, and M.~Maurer,
  ``Towards a skill- and ability-based development process for self-aware
  automated road vehicles,'' in \emph{IEEE ITSC 2017}.\hskip 1em plus 0.5em
  minus 0.4em\relax Piscataway, NJ: IEEE, 2017, pp. 1--6.

\bibitem{Czarnecki.2018b}
K.~Czarnecki, ``Operational world model ontology for automated driving systems
  - part 2: Road users, animals, other obstacles, and environmental
  conditions.''

\bibitem{Glatzki.91920219222021}
F.~Glatzki, M.~Lippert, and H.~Winner, ``Behavioral attributes for a
  behavior-semantic scenery description (bssd) for the development of automated
  driving functions,'' in \emph{2021 IEEE International Intelligent
  Transportation Systems Conference (ITSC)}.\hskip 1em plus 0.5em minus
  0.4em\relax IEEE, 9/19/2021 - 9/22/2021, pp. 667--672.

\bibitem{Popper.1959}
K.~R. Popper, \emph{The logic of scientific discovery}, 1959.

\bibitem{Bagschik.2018b}
G.~Bagschik, T.~Menzel, and M.~Maurer, ``Ontology based scene creation for the
  development of automated vehicles,'' in \emph{2018 IEEE Intelligent Vehicles
  Symposium (IV)}.\hskip 1em plus 0.5em minus 0.4em\relax [S.l.]: IEEE, 2018,
  pp. 1813--1820.

\bibitem{Schuldt.2013}
\BIBentryALTinterwordspacing
F.~Schuldt, F.~Saust, B.~Lichte, M.~Maurer, and S.~Scholz, ``{Effiziente
  systematische Testgenerierung für Fahrerassistenzsysteme in virtuellen
  Umgebungen},'' 2013. [Online]. Available:
  \url{https://publikationsserver.tu-braunschweig.de/receive/dbbs_mods_00052570}
\BIBentrySTDinterwordspacing

\bibitem{Bagschik.2018}
G.~Bagschik, T.~Menzel, C.~K{\"o}rner, and M.~Maurer, ``Wissensbasierte
  szenariengenerierung f{\"u}r betriebsszenarien auf deutschen autobahnen,'' in
  \emph{12. Workshop Fahrerassistenzsysteme und automatisiertes Fahren}, 2018.

\bibitem{Scholtes.2021}
M.~Scholtes, L.~Westhofen, L.~R. Turner, K.~Lotto, M.~Schuldes, H.~Weber,
  N.~Wagener, C.~Neurohr, M.~H. Bollmann, F.~Körtke, J.~Hiller, M.~Hoss,
  J.~Bock, and L.~Eckstein, ``6-layer model for a structured description and
  categorization of urban traffic and environment,'' \emph{IEEE Access},
  vol.~9, pp. 59\,131--59\,147, 2021.

\bibitem{Ulbrich.2014}
S.~Ulbrich, T.~Nothdurft, M.~Maurer, and P.~Hecker, ``Graph-based context
  representation, environment modeling and information aggregation for
  automated driving,'' in \emph{Intelligent Vehicles Symposium Proceedings,
  2014 IEEE}.\hskip 1em plus 0.5em minus 0.4em\relax IEEE, 2014, pp. 541--547.

\bibitem{Buechel.2017}
M.~Buechel, G.~Hinz, F.~Ruehl, H.~Schroth, C.~Gyoeri, and A.~Knoll,
  ``Ontology-based traffic scene modeling, traffic regulations dependent
  situational awareness and decision-making for automated vehicles,'' in
  \emph{28th IEEE Intelligent Vehicles Symposium}, I.~I.~V. Symposium,
  Ed.\hskip 1em plus 0.5em minus 0.4em\relax [Piscataway, NJ]: IEEE, 2017, pp.
  1471--1476.

\bibitem{Hulsen.2011}
M.~Hulsen, J.~M. Zollner, and C.~Weiss, ``Traffic intersection situation
  description ontology for advanced driver assistance,'' in \emph{2011 IEEE
  Intelligent Vehicles Symposium (IV 2011)}.\hskip 1em plus 0.5em minus
  0.4em\relax Piscataway, NJ: IEEE, 2011, pp. 993--999.

\bibitem{Regele.2008}
R.~Regele, ``Using ontology-based traffic models for more efficient decision
  making of autonomous vehicles,'' in \emph{Fourth International Conference on
  Autonomic and Autonomous Systems, 2008}, D.~Greenwood, Ed.\hskip 1em plus
  0.5em minus 0.4em\relax Los Alamitos, Calif. [u.a.]: {IEEE Computer Society},
  2008, pp. 94--99.

\bibitem{Butz.2020}
M.~Butz, C.~Heinzemann, M.~Herrmann, J.~Oehlerking, M.~Rittel, N.~Schalm, and
  D.~Ziegenbein, ``Soca: Domain analysis for highly automated driving
  systems,'' in \emph{2020 IEEE 23rd International Conference on Intelligent
  Transportation Systems (ITSC)}.\hskip 1em plus 0.5em minus 0.4em\relax IEEE,
  2020, pp. 1--6.

\bibitem{.2020}
\BIBentryALTinterwordspacing
``{ASAM OpenDRIVE: Open Dynamic Road Information for Vehicle Environment},''
  2020. [Online]. Available:
  \url{https://www.asam.net/standards/detail/opendrive/}
\BIBentrySTDinterwordspacing

\bibitem{Bender.2014}
P.~Bender, J.~Ziegler, and C.~Stiller, ``Lanelets: Efficient map representation
  for autonomous driving,'' in \emph{Intelligent Vehicles Symposium
  Proceedings, 2014 IEEE}.\hskip 1em plus 0.5em minus 0.4em\relax IEEE, 2014,
  pp. 420--425.

\bibitem{Censi.2019}
\BIBentryALTinterwordspacing
A.~Censi, K.~Slutsky, T.~Wongpiromsarn, D.~Yershov, S.~Pendleton, J.~Fu, and
  E.~Frazzoli, ``Liability, ethics, and culture-aware behavior specification
  using rulebooks.'' [Online]. Available:
  \url{https://arxiv.org/pdf/1902.09355}
\BIBentrySTDinterwordspacing

\bibitem{Rizaldi.2017}
A.~Rizaldi, J.~Keinholz, M.~Huber, J.~Feldle, F.~Immler, M.~Althoff,
  E.~Hilgendorf, and T.~Nipkow, ``Formalising and monitoring traffic rules for
  autonomous vehicles in isabelle/hol,'' in \emph{Integrated Formal Methods},
  ser. Lecture Notes in Computer Science, N.~Polikarpova and S.~Schneider,
  Eds.\hskip 1em plus 0.5em minus 0.4em\relax Cham: {Springer International
  Publishing}, 2017, vol. 10510, pp. 50--66.

\bibitem{Esterle.18.11.202016.12.2020}
K.~Esterle, L.~Gressenbuch, and A.~Knoll, ``Formalizing traffic rules for
  machine interpretability,'' in \emph{2020 IEEE 3rd Connected and Automated
  Vehicles Symposium (CAVS)}.\hskip 1em plus 0.5em minus 0.4em\relax IEEE,
  18.11.2020 - 16.12.2020, pp. 1--7.

\bibitem{Stolte.2017}
T.~Stolte, G.~Bagschik, A.~Reschka, and M.~Maurer, ``Hazard analysis and risk
  assessment for an automated unmanned protective vehicle,'' in \emph{28th IEEE
  Intelligent Vehicles Symposium}, I.~I.~V. Symposium, Ed.\hskip 1em plus 0.5em
  minus 0.4em\relax [Piscataway, NJ]: IEEE, 2017, pp. 1848--1855.

\bibitem{Thorn.2018}
E.~Thorn, S.~Kimmel, and M.~Chaka, ``A~framework for automated driving system
  testable cases and scenarios,'' Washington, D.C.

\bibitem{Waymo.2020}
Waymo, ``Waymo safety report.''

\bibitem{PerdomoLopez.2017}
D.~{Perdomo Lopez}, R.~Waldmann, C.~Joerdens, and R.~Rojas, ``Scenario
  interpretation based on primary situations for automatic turning at urban
  intersections,'' in \emph{VEHITS 2017}, O.~Gusikhin, M.~Helfert, and
  A.~Pascoal, Eds.\hskip 1em plus 0.5em minus 0.4em\relax Set{\'u}bal:
  {SCITEPRESS - Science and Technology Publications Lda}, 2017, pp. 15--23.

\bibitem{Pek.2020}
C.~Pek, S.~Manzinger, M.~Koschi, and M.~Althoff, ``Using online verification to
  prevent autonomous vehicles from causing accidents,'' \emph{Nature Machine
  Intelligence}, vol.~2, no.~9, pp. 518--528, 2020.

\bibitem{JOSM.2021}
\BIBentryALTinterwordspacing
``{JOSM: extensible editor for OpenStreetMap},'' 2021. [Online]. Available:
  \url{https://josm.openstreetmap.de/}
\BIBentrySTDinterwordspacing

\bibitem{Haklay.2008}
M.~Haklay and P.~Weber, ``Openstreetmap: User-generated street maps,''
  \emph{IEEE Pervasive Computing}, vol.~7, no.~4, pp. 12--18, 2008.

\bibitem{Woopen.2018}
\BIBentryALTinterwordspacing
T.~Woopen, L.~Eckstein, S.~Kowalewski, D.~Moormann, M.~Maurer, R.~Ernst,
  H.~Winner, S.~Katzenbeisser, M.~Becker, C.~Stiller, K.~Furmans, K.~Bengler,
  M.~Lienkamp, H.-C. Reuss, K.~Dietmayer, H.~Lategahn, N.~Siepenk\"{o}tter,
  M.~Elbs, E.~V.~Hin\"{u}ber, M.~Dupuis, and C.~Hecker,
  ``\BIBforeignlanguage{en}{{UNICAR}agil - disruptive modular architectures for
  agile, automated vehicle concepts},'' \emph{\BIBforeignlanguage{en}{Volume 1
  / 2018 27th Aachen Colloquium Automobile and Engine Technology 2018}}, vol.
  Aachen, pp. 8 Oct 2018--10 Oct 2018; Aachen : Aachener Kolloquium Fahrzeug,
  2018. [Online]. Available:
  \url{http://publications.rwth-aachen.de/record/749158}
\BIBentrySTDinterwordspacing

\end{thebibliography}
